\documentclass[aps,prl,showpacs]{revtex4}

\usepackage{psfig} \usepackage{epsfig} \usepackage{amssymb}
\usepackage{amsmath}

  \def\Vls{\overline{V}} 
\def\Vss{\delta V} \def\R{\mathbb{R}} 
 \def\<{\langle} \def\>{\rangle}
\begin{document}

\title{Probing Multi-Scale Energy Landscapes Using the String Method}

\author{Weinan E$^1$, Weiqing Ren$^2$, and Eric Vanden-Eijnden$^2$}
\affiliation{$^1$Department of Mathematics and PACM,
  Princeton University, Princeton, NJ 08544\\
  $^2$Courant Institute, New York University, New York, NY 10012 }


\begin{abstract}
  A novel and powerful method is presented for the study of rare
  switching events in complex systems with multiscale energy
  landscapes.  The method performs an umbrella sampling of the
  equilibrium distribution of the system in hyperplanes normal to the
  effective transition pathway, i.e. the transition pathway of a
  coarse-grained potential which need not to be known beforehand.
  Rather the effective transition pathway is determined on the fly in
  an adaptive way by evolving a smooth curve, called a string, that
  connects the initial and final regions and feels thermally averaged
  potential forces.  Appropriate averages around the string determine
  the transition rates.  Application to an example of solid-solid
  transformation of a condensed system is presented.
\end{abstract}

\maketitle

\twocolumngrid

The evolution of complex systems often involves widely separated time
scales.  Well-known examples include nucleation events during phase
transition, conformational changes of macromolecules, and chemical
reactions. The separation of time scale is a consequence of the
disparity between the effective thermal energy and typical energy
barrier of the systems. The state of the system is confined for long
periods of time in metastable regions in configuration space and only
rarely switches from one region to another. The dynamics of the system
effectively reduces to a Markov chain on the metastables regions.
Finding the transition rates between the metastable states is a major
computational challenge.  For low dimensional systems, pathways and
rates can be determined by identifying the minimal energy paths (MEPs)
which connect different minima on the energy surface by going through
saddle points, and the curvature around these paths. For higher
dimensional systems, such a procedure becomes impractical since the
energy surfaces is typically dense in critical points and a multitude
of dynamic paths contribute to the transition.  The transition
pathways have to be understood in an averaged sense as the
superposition of the many paths by which transition between two
metastable regions in phase space takes place.

Sophisticated numerical techniques have been developed for finding
transition pathways and rates \cite{mep0,NEB,TPS}, though most methods
apply when the energy landscape is relatively smooth (e.g. when
transition is accomplished by a few isolated MEPs), and become less
useful otherwise. A notable exception is the transition path sampling
technique (TPS) \cite{TPS} which applies to situations when the
potentials are multiscaled.  TPS samples the probability of dynamical
trajectories conditional on their end points being in given regions of
the configuration space. The transition rates are determined by
reconstructing the unconditional probability of the trajectories
through umbrella sampling of trajectories whose end points are
constrained in overlapping sub-regions covering configuration space.
TPS computes rates directly (i.e. without relating them to the
geometry of the potential).  In this Letter, we propose an alternative
approach which performs an umbrella sampling of the equilibrium
distribution of the system in hyperplanes normal to the transition
pathways of a coarse-grained potential which need not be determined
beforehand.  Appropriate averages then determine the transition rates.

We shall focus on the example of a system modeled by
\begin{equation}
  \label{eq:highfriction}
  \gamma \dot q = -\nabla V(q) + \xi(t)
\end{equation}
where $\gamma$ is the friction coefficient and $\xi(t)$ is a white
noise with $\< \xi_j(t) \xi_k(0)\> = 2 \gamma k_B T \delta_{jk}
\delta(t)$.  In molecular dynamics, (\ref{eq:highfriction}) is the
high friction (or Smoluchowski) limit, but the method can be
generalized easily to arbitrary friction.  We assume that the
potential $V(q)$ is multi-scaled in the sense that there is a gap
between $k_B T$ and the relevant energy barriers of the system.  The
system may certainly have other energy barriers that are comparable or
smaller than $k_B T$.  In this case, even though $V$ may contain a
very large number of local minima, the system experiences a much
smoother energy landscape because of thermal effects and we are
interested in the ``thermally averaged'' energy landscape: $\Vls:= \<
V\>$. The dynamics in both $V$ and $\Vls$ reduce to the same Markov
chain on suitable metastable basins separated by the large energy
barriers. In principle, one can study the transitions in such systems
by first computing $\Vls$ and then applying methods for smooth energy
landscapes such as the nudged elastic band (NEB) method \cite{NEB} or
the zero-temperature string method \cite{ereva02} to study the
transition in the system with potential $\Vls$.  But in practice it is
often impossible to first find $\Vls$.  The motivation of the present
paper is to develop a method that is based directly on $V$ but
computes the {\it effective} transition pathways and rates associated
with $\Vls$.  This is accomplished by a simple but important
modification of the string method introduced in \cite{ereva02} by
incorporating thermal averages.  The thermal average can either be the
result of finite temperature, or the effect of noise introduced
numerically to smooth out the small scale features of the energy
landscape.

It is useful to first review the zero-temperature string method
\cite{ereva02}.  Let $\varphi$ be a smooth curve, which we call a
string, connecting two minima of $V$, $A$ and $B$.  By definition,
$\varphi$ is a MEP if
\begin{equation}
  \label{eq:MEP}
  0=(\nabla V({\varphi}))^\perp,
\end{equation}
where $(\nabla V)^\perp = \nabla V - (\nabla V \cdot \hat t) \hat t$
and $\hat t$ is the unit tangent vector along $\varphi$, $\hat t =
\varphi_\alpha/|\varphi_\alpha|$.  Equivalently $\varphi$ is a curve
which minimizes $V$ in the hyperplane normal to itself.  We will use
suitable parametrizations of the string such that $\alpha=0$ at $A$,
$\alpha=1$ at $B$.  One way of finding solutions of (\ref{eq:MEP}) is
to follow the dynamics defined by:
\begin{equation}
  \label{eq:string}
  \varphi_t = -(\nabla V (\varphi))^\perp+ r \hat t.
\end{equation}
where for convenience we renormalized time as $t\to t/\gamma$.  The
scalar field $r\equiv r(\alpha,t)$ is a Lagrange multiplier term added
in (\ref{eq:string}) to preserve some constraint on the
parametrization of $\varphi$.  For instance, a simple choice is to
impose that $\varphi$ be parametrized by normalized arclength.  In
this case (\ref{eq:string}) must be supplemented by the constraint $
(|\varphi_\alpha|)_\alpha =0$ which determines $r$.  Other
parametrizations -- for instance by energy-weighted arclength -- can
be straightforwardly implemented by modifying this constraint.  See
\cite{ereva02} for details.

This method is very effective if the energy landscape is smooth, as
shown in \cite{ereva02}.  In this case, it bears a lot of similarity
with NEB \cite{NEB}. But as we now show, it is the differences between
the string method and NEB, namely the use of continuous curves with
fixed and intrinsic parametrization, that allows us to extend the
string method to the case of rough energy surfaces.

When the energy landscape is multiscaled, our objective is to compute
the effective transition pathways (MEPs in $\Vls$) and rates, without
computing first $\Vls$.  (Let us emphasis that since $\Vls$ is
relatively smooth and the energy barriers associated with $\Vls$ are
much larger than $k_B T$, we can assume that the effective MEPs are
isolated.)  To this end, we make an important modification of
(\ref{eq:string}), namely we replace the gradient of $V$ at the right
hand side of (\ref{eq:string}) by an ensemble and/or temporal average
of the actual force among a number of string configurations. We denote
this ensemble by $\{\varphi^\omega\}$, its mean by $\varphi^\circ:= \<
\varphi^\omega\>$, and therefore replace (\ref{eq:string}) by
\begin{equation}
  \label{eq:meanstring}
  \varphi^\circ_t = -\< \nabla V (\varphi^\omega)\> ^{\perp,\circ}
  + r^\circ \hat t^\circ,
\end{equation}
where $\hat t^\circ$ is the unit tangent vector along $\varphi^\circ$
and $(\cdot)^{\perp,\circ}$ denotes the projection to the hyperplane
normal to $\hat t^\circ$. To preserve parametrization of
$\varphi^\circ$, (\ref{eq:meanstring}) must be supplemented by some
appropriate constraint which determines $r^\circ$: for instance, the
constraint $(|\varphi^\circ_\alpha|)_\alpha =0$ corresponds to
normalized arc-length.

A practical way to create the ensemble $\{\varphi^\omega\}$ is to
introduce a finite temperature version of (\ref{eq:string}) through
\begin{equation}
  \label{eq:nstring1}
  \varphi^\omega_t = -(\nabla V(\varphi^\omega)) ^{\perp, \circ} 
  + r^\circ \hat t^\circ+  
  (\eta^\omega)^{\perp, \circ},
\end{equation}
where $\eta^\omega$ is a noise term added in order
to simulate the effect of thermal averaging.
We take $\eta^\omega$ to be a Gaussian process with covariance
\begin{equation}
  \label{eq:noisedef}
  \<\eta^\omega_j (\alpha,t) \eta^\omega_k (\alpha',0)\> = 
  \begin{cases}2 k_B T \delta_{jk}\delta(t) & 
    \hbox{if  } \alpha =\alpha',\\
    0& \hbox{otherwise},
  \end{cases}
\end{equation}
and compute all averages with respect to the statistics of
$\eta^\omega$. In particular, the mean of (\ref{eq:nstring1}) is
(\ref{eq:meanstring}).

The equilibrium density function for (\ref{eq:nstring1}) is given by
%
\begin{equation}
  \label{eq:IM}
  \mu(q,\alpha) = Z^{-1}(\alpha) e^{-\beta V(q)} 
  \delta_{S^\circ(\alpha)}(q).
\end{equation}
Here $\beta=1/k_B T$, $S^\circ(\alpha)$ is the hyperplane normal to
$\varphi^\circ(\alpha)$, $\delta_{S^\circ(\alpha)}(q)$ is the Dirac
distribution concentrated on $S^\circ(\alpha)$, and
\begin{equation}
  \label{eq:Zdef}
  Z(\alpha) = \int_{S^\circ(\alpha)} e^{-\beta V(q)}d^dq
\end{equation}
is the partition function. (\ref{eq:IM}) can be determined by
analyzing the Fokker Planck equation associated with
(\ref{eq:nstring1}).  The mean string $\varphi^\circ(\alpha)$ entering
(\ref{eq:IM}) through $S^\circ(\alpha)$ has to be determined
self-consistently from
\begin{equation}
  \label{eq:meanstringeq}
  \varphi^\circ(\alpha)= \int_{\R^d} q \mu(q,\alpha)d^dq.
\end{equation}
This equation implies that $\varphi^\circ$ is a MEP for $\Vls$ to
leading order in $k_B T$. Indeed if $V = \Vls + \Vss$ with $\Vss$
comparable to $k_B T$ or smaller, the integral in
(\ref{eq:meanstringeq}) can be evaluated by Laplace method and to
leading order in $k_B T$, only $\Vls$ contributes.  This shows that
$\varphi^\circ$ is the minimum of $\Vls(q)$ in the hyperplane
perpendicular to $\varphi^\circ$.

(\ref{eq:IM}) shows that the stochastic string performs an umbrella
sampling of the equilibrium distribution of the system in the
one-parameter family of hyperplanes normal to a MEP of $\Vls$. (We
stress again that this MEP needs not be known beforehand but rather is
determined on the fly by the method.)  Averages with respect to
(\ref{eq:IM}) yield relevant quantities like, in particular, the
transition rates between the basins of metastability visited by the
mean string.  This follows from a straightforward generalization of
Kramers argument (see e.g. chap.~9 in \cite{gar85}) to multiscale
energy landscapes. Let $ F(\alpha) = -k_BT\ln Z(\alpha)$ be the free
energy. Using the identity $\int \partial \ln Z /\partial \alpha d
\alpha = \ln Z(\alpha)$, we obtain from (\ref{eq:Zdef}) after
integration by part
\begin{equation}
  \label{eq:freeEdiff2}
  F(\alpha) = \int
  \bigl\< (\hat t^\circ \cdot\nabla V) 
    \left((\hat t^\circ \cdot \varphi^\circ)_\alpha
      -\hat t^\circ _\alpha \cdot \varphi\right)\bigr\>d\alpha.
\end{equation}
Let $\alpha_1$, $\alpha_2$, ... be the successive values of $\alpha$
where $F$ reaches a local minimum. Each $\alpha_j$ corresponds to a
basin of metastability visited by the mean string.  Let $\alpha_{12}$,
$\alpha_{23}$, ... be the successive values of $\alpha$ where $F$
reaches a local maximum. Then
\begin{displaymath}
  k^{-1}_{1\to 2} = 2 \gamma \beta \int_{\alpha_1}^{\alpha_{12}} 
  e^{-\beta F(\alpha)} |\varphi^\circ_\alpha|d\alpha
  \int_{\alpha_1}^{\alpha_2} e^{\beta F(\alpha)} |\varphi^\circ_\alpha|
  d\alpha,
\end{displaymath}
and so on. To leading order in $k_BT$, this expression is equivalent
to
\begin{equation}
  \label{eq:ratehf}
  k_{1\to 2} = \frac{\sqrt{F_{\alpha\alpha}(\alpha_1)
      |F_{\alpha\alpha}(\alpha_{12})|}}
  {2\pi \gamma|\varphi^\circ_\alpha(\alpha_1)|
    |\varphi^\circ_\alpha(\alpha_{12})|} e^{-\beta\mathit{\Delta} F_{12}}
\end{equation}
where $\mathit{\Delta} F_{12}:=F(\alpha_{12}) -F(\alpha_1)$ is the free
energy barrier from basin $1$ to basin $2$.

In practice, (\ref{eq:nstring1}) is solved by considering $M$
realizations of the string, $\varphi^j$, $j=1,\ldots,M$, and
approximating the mean string as $\varphi^\circ = M^{-1} \sum_{j=1}^M
\varphi^j$.  After a number of steps, a reparametrization step for
$\varphi^\circ$ is applied.  Once $\varphi^\circ$ has converged to its
steady state value, no reparametrization step is necessary anymore
and, using ergodicity, one can supplement the ensemble average by a
time average using $\varphi^\circ = (MT)^{-1} \sum_{j=1}^M \int_0^T
\varphi^j(t)dt$ to obtain better statistics. Other averages like
(\ref{eq:freeEdiff2}) are evaluated similarly.


\begin{figure}[t]
  \center \includegraphics[width=7cm]{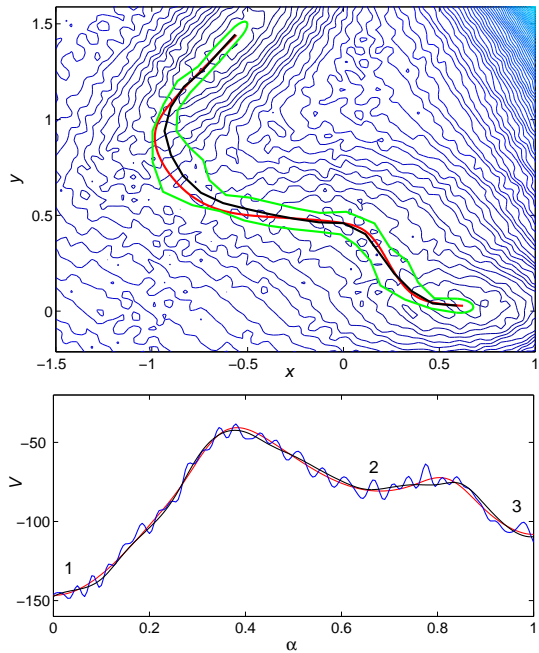} 
  \caption{\label{fig:mueller} 
    Upper panel: the mean transition path $\varphi^\circ$ (black
    curve) on the disordered Mueller potential isolines. Also shown is
    the MEP, $\varphi^\star$, in $\bar V$ (red curve) obtained by
    integrating (\ref{eq:string}).  Both curves coincide fairly well
    as expected.  The region within the green curve is the root mean
    square displacement of the stochastic string indicating the
    typical size of the fluctuations of the string.  Lower panel: the
    mean energy along the stochastic string (black curve), $\< V
    (\varphi)\>$, compared with the smoothed energy along the MEP of
    $\bar V$, $\bar V (\varphi^\star)$ (red curve). The blue curve is
    the total energy along $\varphi^\circ$, $V(\varphi^\circ)$, which
    shows the large number of (irrelevant) minima and saddle points
    crossed during a transition. We took 21 discretization points
    along the stochastic string and 20 realizations. }
\end{figure}

As a first illustrative example of the procedure, we consider a
disordered version of Mueller potential.  We take $\Vls$ to be the
Mueller potential, originally invented as a nontrivial test for
reaction path algorithms \cite{mue80}, and $\Vss$ as
\begin{equation}
  \label{eq:deltaVmueller}
  \begin{array}{r}
    \displaystyle \delta V(x,y) = \sum_{k_1,k_2=5}^{10} 
    \bigl(\alpha_k \cos ( 2 \pi (k_1 x +k_2 y))\qquad\\[-6pt]
    \displaystyle
    +\beta_k \cos ( 2 \pi (k_1 x -k_2 y))\bigr),
  \end{array}
\end{equation}
where $\alpha_k,\beta_k\in[-1,1]$. With this choice $|\delta V| \le
10$ (compared to a main energy barrier of about $100$ in $\bar V$) and
we shall assume that $k_B T=5$.  The number of MEPs in $V = \bar V +
\delta V$ joining the regions of metastability around the two deeper
minima of $\bar V$ is of the order of $10^2$. This makes the complete
sampling of the MEPs in $V$ (say, using (\ref{eq:string}) with the
full $V$, or NEB) impractical (and irrelevant).  The results of the
stochastic string method are summarized in figure~\ref{fig:mueller}.
The MEP in $\bar V$ is successfully retrieved and, using
(\ref{eq:freeEdiff2}), we also obtained the value $\mathit{\Delta}
F_{12} = 109.0$ for the free energy barrier between 1 and 2, compared
to the exact value of $\mathit{\Delta} F_{12} = 108.9$
\cite{rem_muel}.

Our next example illustrates the complex energy landscapes for systems
in condensed phases. We consider a crystal of $20 \times 20$ array of
atoms interacting via a pairwise potential that has three minima at
$r_1 =1, r_2 = 1.1$ and $r_3 = \sqrt{1+1.1^2}$. There are two basic
equilibrium states for this systems, with the position of the
$(i,j)$-th atom given by $(r_1 i, r_2 j)$ and $(r_2 i, r_1 j)$
respectively. Their rigid body rotations are considered to be
equivalent states. The two states can be considered as the two
variants of the martensitic phase. We are interested in the energy
landscape that the system experiences when transforming from one
variant to the other.

For clarity we initiate the seed of transformation at the upper-left
corner. Free boundary conditions are used in the simulation. The
crystal transforms through the propagation of the twin boundary,
oriented diagonally as shown in Figure 2.  The twin boundary
propagates through kink propagation along the twin boundary
associated with the phase transformation of individual atoms.  The
energy landscape exhibits three scales: The largest scale associated
with the position of the twin boundary, an intermediate scale
associated with the propagation of the twin boundary by one atomic
distance, and a small scale associated with the kink propagation.  For
the stochastic string method, we choose $k_B T$ to be in between the
energy barrier for kink propagation and twin boundary propagation.
The energy landscape for the twin boundary to advance one atomic
distance is shown in Figure 2. 40 points are used. In Figure 2 we also
show the potential energy landscape associated with the MEP computed
using the deterministic string method with 400 points. Even though the
details of the energy landscape is not resolved, as expected, the
overall feature is captured very well by the stochastic string method
using only 40 points.

\begin{figure}[t]
  \center \ \includegraphics[width=8.0cm]{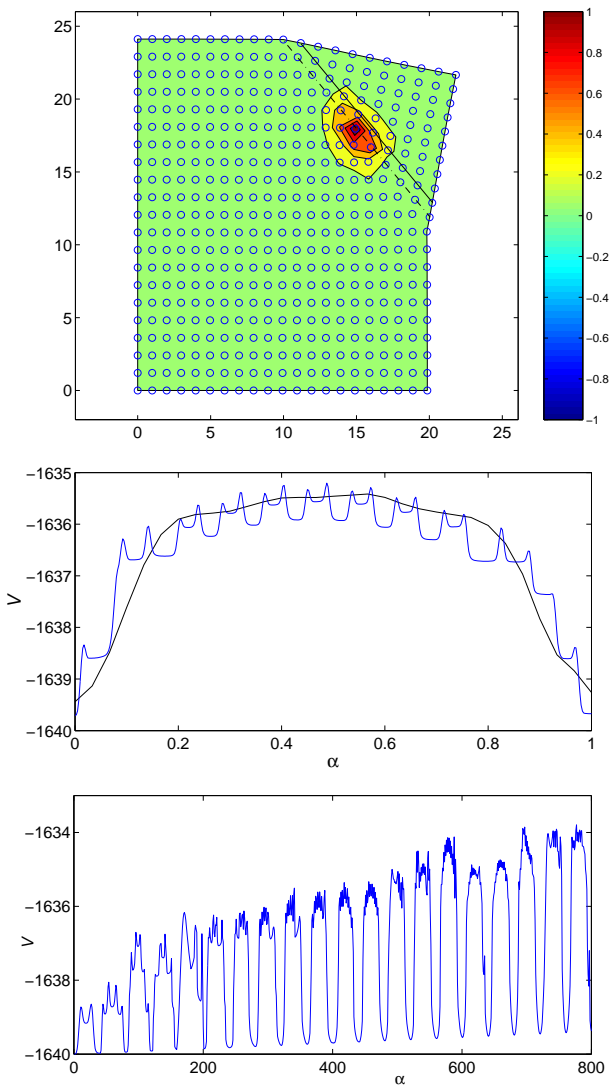}
  \caption{\label{fig:mart} 
    Upper panel: Snapshot of the crystal during transformation.  The
    color bar shows the scale of the local energy of each atom.
    Middle panel: The mean potential, $\< V(\varphi)\>$, experienced
    by the crystal at finite temperature when the twin boundary moves
    by one atomic distance along the diagonal.  This result was
    obtained using $40$ discretization points along the strings.  Also
    shown for comparison is the potential energy along one particular
    MEP (blue curve, at zero temperature) which shows the features of
    the potential associated with the transformation of individual
    atoms along the twin boundary.  $400$ discretization points along
    the string were necessary to get this fully resolved result.
    Lower panel: The energy landscape computed by the string method at
    finite temperature after the crystal is half transformed.  Notice
    the appearance of three scales on the energy landscape.  }
\end{figure}

In conclusion, we have introduced a powerful method to probe
multiscale energy landscape and thereby determine effective transition
pathways, free energy barriers, and transition rates in complex
systems. The method performs umbrella sampling in the hyperplanes
perpendicular to an effective transition pathway which is determined
on the fly in an adaptive way. The method can be further improved in
various ways. For instance the temperature of the string can be
periodically changed on an annealing schedule. This allows to further
explore the energy landscape and determine if more than one effective
transition pathway is involved in a given transition. We are also
investigating the possibility of using Metropolis Monte-Carlo scheme
instead of the Langevin equation~(\ref{eq:nstring1}) to compute
averages with respect to the equilibrium distribution in
(\ref{eq:IM}). This speeds up convergence since it requires to compute
$V$ only instead of its gradient. Finally, in its present form the
method performs the coarse-graining in an implicit way, i.e.  no
reaction coordinate has to be known beforehand. In certain systems
where reaction coordinates can be easily determined, the method can be
applied to these coarse-grained variables alone, which might
dramatically speed up convergence if the reaction coordinates are low
dimensional.

We thank David Chandler and Bob Kohn for helpful discussions.  The
work of E is supported in part by NSF grant DMS01-30107.


\begin{thebibliography}{99}
\bibitem{rem_muel} The exact value for the free energy barrier is
  obtained by determining the MEP in the smooth Mueller potential
  $\Vls$, the energy barrier along this MEP, and the determinant of
  the Hessian at the minima and saddle points.
\end{thebibliography}
\end{document}